\documentstyle[prl,preprint,aps]{revtex}
\begin{document}
%\draft
\title{Stationary Regime of Random Resistor Networks 
\\Under Biased Percolation}
\author{C. Pennetta$^{1}$, L. Reggiani$^{1}$, E. Alfinito$^1$ 
and G. Trefan$^2$}
\address{$^1$ INFM - National Nanotechnology Laboratory and 
Dipartimento di Ingegneria dell'Innovazione, \\Universit\`a di Lecce,
Via Arnesano, I-73100 Lecce, Italy\\ $^2$ Department of Electrical
Engineering, Eindhoven University of Technology, \\5600 MB Eindhoven,
The Netherlands
\thanks{Corresponding authors e-mail: cecilia.pennetta@unile.it}}
%\date{\today}
\maketitle
\vspace{-0.5 cm}
\begin{abstract}
\noindent
The state of a 2-D random resistor network, resulting from 
the simultaneous evolutions of two competing biased percolations, is 
studied in a wide range of bias values. Monte Carlo simulations show 
that when the external current $I$ is below the threshold value  
for electrical breakdown, the network reaches a steady state with a nonlinear 
current-voltage characteristic. The properties of this nonlinear regime 
are investigated as a function of different model parameters. 
A scaling relation is found between $<R>/<R>_0$ and $I/I_0$, where $<R>$ is 
the average resistance, $<R>_0$  the linear regime resistance and 
$I_0$ the threshold value for the onset of nonlinearity. 
The scaling exponent is found to be independent of the model parameters. 
A similar scaling behavior is also found for the relative variance of 
resistance fluctuations. These results compare well with resistance 
measurements in composite materials performed in the Joule regime up to 
breakdown.

\end{abstract}


%\begin{multicols} {2}

%\narrowtext

\section{Introduction and model}

Electrical breakdown of disordered media has been widely 
studied in  the last twenty years \cite{hermann}-\cite{pen_prl_fail}. 
This is due to its relevant implications on  material technology and on   
 fundamental aspects related to the understanding of the response properties
of disordered systems to high external stresses 
\cite{hermann,bardhan}. 
It is well known that the application of a finite stress 
(electrical or mechanical) to a disordered material generally implies a 
nonlinear response, which leads to a catastrophic behavior in the high stress 
limit \cite{hermann,bardhan}. 
Percolation theory provides a powerful approach for studying breakdown 
phenomena of disordered media \cite{stauffer,stanley}. In this framework, 
several models have been proposed to  describe the electrical breakdown of 
granular metals and of conductor-insulator composites in terms of critical 
phenomena near the percolation threshold \cite{duxbury}-\cite{sornette96}.
Furthermore, the critical exponents characterizing these behaviors 
have been both theoretically calculated and measured in severals 
materials \cite{hermann,bardhan},\cite{duxbury}-\cite{sornette96},
\cite{pen_prl_fail}. Nevertheless, few attempts have been made so far to 
describe the behavior of disordered media over the full range of applied 
stress \cite{sornette96,mukherjee}. 
The understanding of breakdown phenomena in the full dynamical regime 
is thus unsatisfactory to the present time. On the other hand, important 
information is expected from such a study, like precursor effects, r\^ole of 
disorder, predictability of breakdown, etc. \cite{sornette96,mukherjee}.

The aim of this paper is to present a percolative model of sufficient 
generality to address the above issues. To this purpose, we consider a random 
resistor network (RRN) \cite{stauffer} in which two competing percolations 
are present, defect generation and defect recovery, which determine the 
values of the elementary network resistances. Both processes are driven by 
an external current and by the heat exchange between the network and the 
thermal bath. Monte Carlo simulations are performed to explore the network 
evolution in a wide range of bias values. A stationary state or an 
irreversible breakdown of the RRN can be reached depending on the value of 
the applied current. By focusing on the steady state, we study the resistance 
and the resistance noise properties. 
We found that the average network resistance 
and the relative resistance noise scale with 
the ratio of the applied current to the current value corresponding to the 
onset of nonlinearity. These results are  discussed in connection with  
measurements in composite materials and in conducting polymers 
\cite{mukherjee,nandi}.

We consider a two-dimensional, square-lattice RRN of total resistance $R$, 
made of $N_{tot}$ resistors with resistance $r_{n}$. We take a square geometry 
$N\times N$, where $N$ determines the linear size of the network.
A constant current $I$ is applied through  electrical contacts realized 
by perfectly conducting bars at the left and right sides of the network. As a
consequence, a current $i_n$ is flowing in the n{\em th} resistor. 
The RRN interacts with a thermal bath at  temperature $T_0$ and the 
resistances $r_{n}$ are taken to depend linearly on the local temperature 
$T_n$, according to:  
	\begin{equation}
	r_{n}(T_{n})=r_{0}[ 1 + \alpha (T_{n} - T_0)]
	\label{eq:tcr}
	\end{equation}
where $r_{0}$ is the resistance value of the elementary resistor 
at the temperature $T_{0}$ and $\alpha$ is the temperature coefficient of 
the resistance. The local temperatures are calculated as in 
Ref. \cite{pen_prl_fail}: 
	\begin{equation}
	T_{n}=T_{0} + A \Bigl[ r_{n} i_{n}^{2} + {B \over N_{neig}}
      \sum_{l=1}^{N_{neig}}  \Bigl( r_{l} i_{l}^2   - r_n i_n^2 \Bigr) \Bigr].
	\label{eq:temp}
	\end{equation}
In this expression, $N_{neig}$ is the number of first neighbours of the 
n{\em th} resistor, the parameter $A$, measured in (K/W), describes the 
heat coupling of each resistor with the thermal bath and it determines the 
importance of Joule heating effects. The parameter $B$ is taken to be equal 
to $3/4$ to provide a uniform heating in the perfect network configuration.
We notice that Eq.~(\ref{eq:temp}) implies an instantaneous thermalization of
each resistor at the value $T_n$, and then, by adopting  Eq.~(\ref{eq:temp}),
we are neglecting for simplicity time dependent effects which are discussed 
in Ref. \cite{sornette92}. 

In the initial state of the network, $I=0$, $T_n \equiv T_0$ and all the 
resistors are identical $r_n \equiv r_0$. The evolution of the RRN arises 
from the presence of two competing percolations. The first consists of 
generating fully insulating defects (broken resistors). This process occurs 
with probability: 
	\begin{equation}
	W_D=\exp[ -E_D/K_B T_n ] 
	\label{eq:prob}
	\end{equation}
where $E_D$ is a characteristic activation energy and $K_B$ 
the Boltzmann constant \cite{pen_prl_fail}. The second percolation consists of 
recovering the insulating defects. This process occurs with a probability 
$W_R$ expressed as in Eq.~(\ref{eq:prob}) but with a different activation 
energy $E_R$. As a result of the competition between these two percolations, 
depending on the parameters which specify the physical properties of the 
system and depending on the external conditions (bias current and  
bath temperature), the RRN can reach a steady state or the percolation
threshold. 
In the first case, $R$ fluctuates around its average value $<R>$, 
while in the second case, an irreversible breakdown occurs, i.e. $R$ diverges 
due to the existence of at least one continuous path of defects between the 
upper and lower sides of the network \cite{stauffer}. By focusing on the 
effect of the external current, we define $I_b$ as the greatest current value 
for which the RNN is stationary. We notice that for biased percolation the 
following expression, $E_R < E_D + k_BT \ln [1 + \exp(-E_D/K_BT_0)]$, 
represents a necessary condition for the existence of a 
steady state \cite{pen_prl_stat,unpub}.
Monte Carlo simulations are performed according to the following procedure: 
(i) starting from the perfect lattice with given local currents and 
temperatures, $i_n$ and $T_n$, (ii) resistances $r_n$ are changed 
according to Eq.~(\ref{eq:tcr}) and  defects are generated with 
probability $W_D$; (iii) the currents $i_n$ are calculated by solving 
Kirchhoff's loop equations; the local temperatures are updated; (iv) 
the temperature dependence of the resistances $r_n$ is again accounted for
and  defects are recovered with probability $W_R$; (v) $R$, $i_n$ and 
$T_n$ are finally calculated and this procedure is iterated from (ii) until 
electrical breakdown or steady state is achieved. In the last case the 
iteration runs long enough to allow a fluctuation analysis to be performed. 
Each iteration step can be associated with an elementary time step on an 
appropriate time scale (to be calibrated with experiments). 
As reasonable values of the parameters, simulations have been performed 
by taking: $N =75$ (except when differently specified), $T_0=300$ (K), 
$\alpha = 10^{-3}$ (K$^{-1}$), $A=5 \times 10^5$ (K/W), $E_D = 0.17$ (eV). 
Several values of $E_R$ and $r_0$ have been considered: 
$0.026 \le E_R \le 0.16$ (eV) and $1 \le r_0 \le 10$ ($\Omega$). 
The values of the external current range between $0.001$ and $3$ (A).  

\section{Results}
We show in Fig. 1 a picture of the RRN near the electrical breakdown.
In this case we have taken: $N=45$, $r_0=1$ ($\Omega$), $E_R= 0.10$ (eV) and 
$I=0.5$ (A), i.e. $I>I_b=0.45$ (A). The different levels of gray correspond 
to different values of $r_n$. We can clearly see that, with respect to
the initial state (perfect network), the network has evolved 
to a disordered state associated with the formation and growth of a channel 
of broken resistors elongated in the direction perpendicular to the applied 
current. This kind of damage pattern well reproduces the experimental 
pattern observed in metallic lines failed as a consequence of 
electromigration \cite{ohring}.
Typical evolutions of $R$ are shown in Fig. 2 at increasing bias values. 
In this case $N=75$ while all the other parameters are the same
of Fig. 1. Thinner curves refer to steady state regime while the thicker curve
refers to a RRN undergoing electrical breakdown ($I>I_b=0.75$).
In the steady state two regimes can be identified: an Ohmic regime (lower
two curves) and a nonlinear regime characterized by a significant increase
of resistance (remaining curves). 
By focusing on the steady state regime, we report in Fig. 3 the 
average resistance $<R>$ as a function of the applied current. The different 
curves correspond to different values of $r_0$, i.e. to RRNs of different 
initial resistance, while the recovery activation energy is $E_R= 0.026$ (eV).
Each value have been calculated by considering the time average on a single 
realization and then making the ensemble average over 20 realizations. 
At low biases the average resistance takes a constant value $<R>_0$ 
which represents 
the intrinsic linear response property of the network (Ohmic regime). 
When $I$ is above a threshold value $I_0$ (threshold for the nonlinearity 
onset), $<R>$ increases with bias until the applied current reaches the $I_b$ 
value, above which the RRN undergoes electrical breakdown. Thus, in the 
following, we indicate with $<R>_b$ the average value of $R$ at $I=I_b$, i.e. 
the last stable value of the resistance. Figure 3 also shows that by increasing
$r_0$ and thus the initial network resistance, 
both $I_0$ and $I_b$ decrease. 
Precisely, we have found: $I_{b} \sim R_{0}^{-\delta}$ and  
$I_{0} \sim R_{0}^{-\delta}$ with $\delta=0.51 \pm 0.01$. Therefore the 
ratio $I_{b}/I_0=28 \pm 1$ is independent of the initial network resistance. 
Moreover, we have also found $<R>_b/<R>_0=1.85 \pm 0.08$.
The effect of the recovery activation energy on the steady state is shown in 
Fig. 4, which reports the ratio $<R>/<R>_0$ as a function 
of the applied current for different values of $E_R$ (in this case all the 
curves are obtained for $r_0=1$ ($\Omega$)). The overall behavior is similar 
to that shown in Fig. 3: an Ohmic regime at low bias is followed by a 
nonlinear regime for $I>I_0$. Moreover, by increasing $E_R$ both $I_0$ and  
$I_b$ decrease and the stability region is thus strongly reduced. 
Nevertheless, an important difference between the effect of varying the 
initial network resistance and that of varying $E_R$ is that, in the last 
case,  the ratio $I_{b}/I_0$, exhibits a significant dependence on $E_R$, 
as shown in Fig. 5.    
To investigate the existence of scaling relations and their universality, 
Fig. 6 reports the log-log plot of the relative variation of the average 
resistance, $(<R>-<R>_{0})/<R>_{0}$, as a function of the ratio $I/I_0$ for 
different values of $r_0$ and $E_R$. The plot shows that all these curves 
collapse onto a single one and that the relative variation of $<R>$ as a 
function of $I/I_0$  exhibits a power law behavior. 
We conclude that, the average resistance follows the scaling relation:    
	\begin{equation}
	\frac{<R>}{<R>_{0}} = g(I/I_{0}), \qquad 
 	g(I/I_{0}) \simeq 1+ (I/I_{0})^{\theta}
	\label{eq:scal}
	\end{equation}
with the scaling exponent $\theta=2.1 \pm 0.1$ being independent of 
both the initial resistance of the RRN and the recovery activation energy. 
Other simulations, performed on rectangular networks, confirm 
the same value for $\theta$. The quadratic dependence of $<R>$ on $I$, 
found here, can be understood in the spirit of mean-field theory when we 
consider that $\Delta R \approx \alpha R_0 \Delta T$ and  
$\Delta T \propto A R_0 I^2$. 
Moreover, recalling previous results concerning the ratio $I/I_{0}$, 
Eq.(\ref{eq:scal}) explains the independence of the ratio $<R>_b/<R>_0$ 
on the initial RRN resistance (Fig. 3) and, by contrast, its significant 
dependence on $E_R$, as shown in Fig. 4. 
All these results well agree with recent measurements in the Joule regime of 
carbon high-density polyethylene composites reported in Ref. \cite{mukherjee}.

The resistance fluctuations are analyzed for different values of $E_{R}$ and 
$r_{0}$. Figure 7 reports the relative variance of resistance fluctuations,
$\Sigma \equiv <\Delta R^2>/<R>^{2}$, as a function of the external 
current. Curves 1, 2 and 3 (with full circles) show $\Sigma$ for $r_{0}=1$ 
($\Omega$) and $E_{R}=0.060$, $0.043$, $0.026$ (eV) respectively, while the 
curves belonging to the set 3 are obtained for $E_{R}=0.026$ (eV) 
and different 
values of $r_{0}$. Figure 7 points out the existence of two different noise 
regimes. The first regime occurs for $I<I_0$, i.e. when Joule heating effects 
are negligible. This noise arises from two random percolations and represents 
an intrinsic noise of the RRN, depending only on the values of $E_D$ and 
$E_R$ \cite{pen_prl_stat}. The second regime occurs when $I>I_0$ and the 
value of $\Sigma$ is found to become strongly dependent on the external 
current. By plotting $\Sigma/\Sigma_0$ as a function of $I/I_{0}$ we have 
found that all the data of Fig. 7 collapse onto a single curve, as shown in 
Fig. 8. Moreover, a power law behavior is observed in the pre-breakdown region.
Therefore, we can conclude that the relative variance of resistance 
fluctuations 
follows the scaling relation:
	\begin{equation}
	\frac{\Sigma}{\Sigma_{0}} = f(I/I_{0}), \qquad 
 	f(I/I_{0}) \simeq 1 + (I/I_{0})^{\eta}
	\label{eq:nois}
	\end{equation}
where the scaling exponent is $\eta =4.1 \pm 0.1$. This value of $\eta$ 
agrees with the values obtained from  electrical noise measurements 
in conducting polymers \cite{nandi}.

In conclusion, we have studied by Monte Carlo simulations the stationary 
regime of RRNs resulting from the simultaneous evolutions of two competing 
percolations. The two percolations consist of generating (recovering) 
fully insulating defects which are driven by an external current and 
by the heat exchange with a thermal bath. We have analyzed the behavior of the
average resistance and of the relative variance of resistance fluctuations 
over a wide range of the applied current and as a function of different model 
parameters. We have found that both these quantities follow a scaling relation
in terms of the ratio between the applied current and the current value
corresponding to the nonlinearity onset. Both scaling exponents are found to 
be independent of the model parameters. These results compare well with 
resistance measurements in composite materials performed in the Joule regime 
up to breakdown \cite{mukherjee} and with noise measurements in conducting 
polymers \cite{nandi}.

\acknowledgments{This research is performed within the STATE project of INFM.
Partial support is also provided by ASI project, contract I/R/056/01.}

\newpage
\noindent FIGURE CAPTIONS
 
\vskip1pc\noindent
Fig. 1 - Network evolution near the electrical breakdown. Different gray 
levels from black to white correspond to increasing values of $r_n$ from 
1 to 3 ($\Omega$).

\vskip1pc\noindent
Fig. 2 - Resistance as a function of time at increasing bias values. 
Thinner curves correspond to steady state regime and they are obtained,
going from  bottom to top, for $I = 0.01$, $0.05$, $0.10$, $0.35$, 
$0.70$, $0.75$ (A). The thicker curve is obtained for $I=0.78$ (A) and 
it corresponds to a RRN undergoing electrical breakdown. 

\vskip1pc\noindent 
Fig. 3 - Normalized average resistance versus external bias. We take 
$E_R=0.026$ (eV) and $E_D=0.167$ (eV), while the value of $r_{0}$ ranges 
between 1 $\Omega$ and 10 $\Omega$.

\vskip1pc\noindent 
Fig. 4 - Normalized average resistance versus external bias. We take 
$r_{0}= 1$ $(\Omega)$, $E_D=0.167$ (eV) while the values 
of $E_R$ range between $0.026$ (eV) and $0.155$ (eV). 

\vskip1pc\noindent
Fig. 5 - Plot of the ratio $I_{b}/I_{0}$ as function of the recovery
activation energy. The curve is a fit with a power law:
$I_{b}/I_{0} \sim E_R^{-0.36}$.

\vskip1pc\noindent 
Fig. 6 - Log-log plot of the relative variation of resistance 
versus $I/I_{0}$. Data shown in this figure are the same of those reported
in Figs. 3 and 4.

\vskip1pc\noindent
Fig. 7 - Relative variance of resistance fluctuations as a function of the 
external bias. Curves 1, 2, 3 refer to $r_{0}=1.0$ ($\Omega$) and 
$E_R=0.060$, $0.043$, $0.026$ eV, respectively. The five curves belonging to 
set 3 are obtained with  $r_{0}=1.0$ ($\Omega$) (full circles), $2.5$ 
($\Omega$) (open squares), $5.0$ ($\Omega$) (full triangles), $7.5$ ($\Omega$)
(open triangles), $10.0$ ($\Omega$) (full diamonds).

\vskip1pc\noindent
Fig. 8 -  Log-log plot of the relative variance of resistance 
fluctuations normalized to the same 
quantity calculated in the linear regime versus $I/I_0$.  
A power law fit is shown in the pre-breakdown regime.      

%\end{multicols}

%
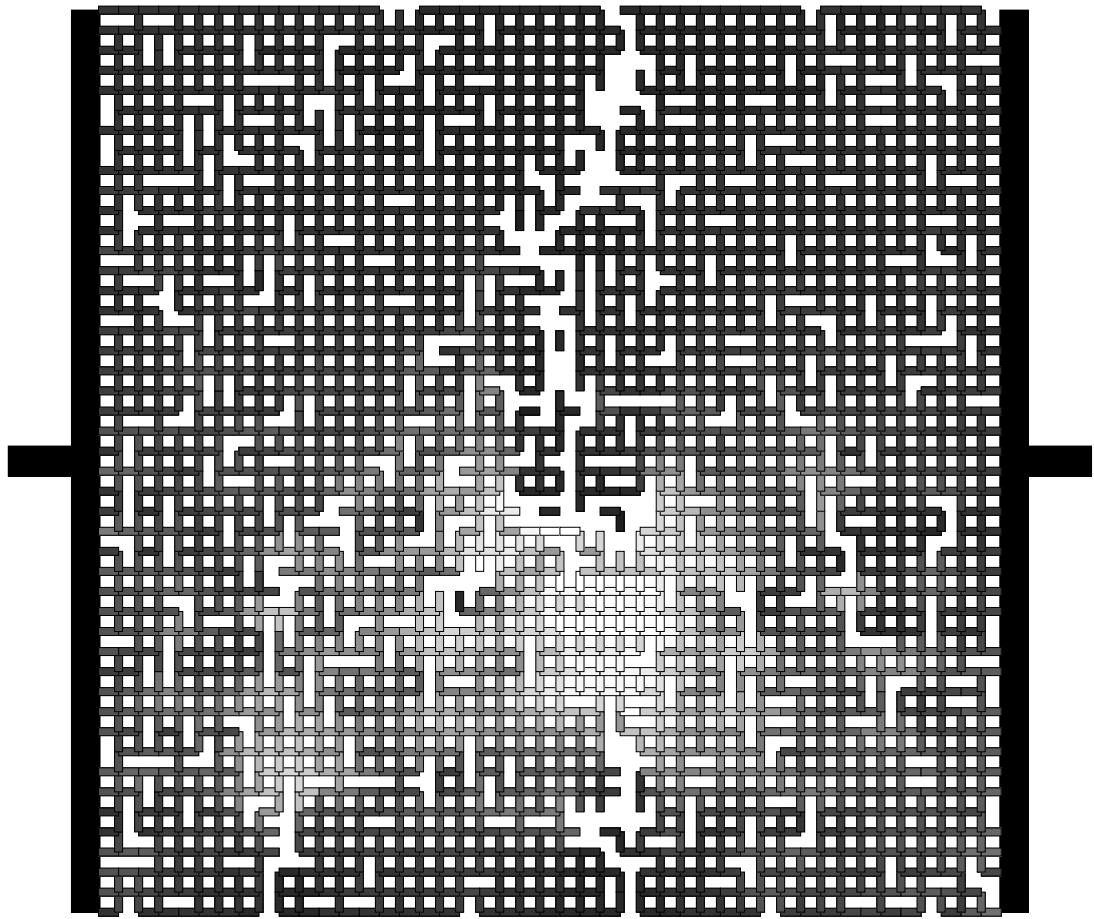
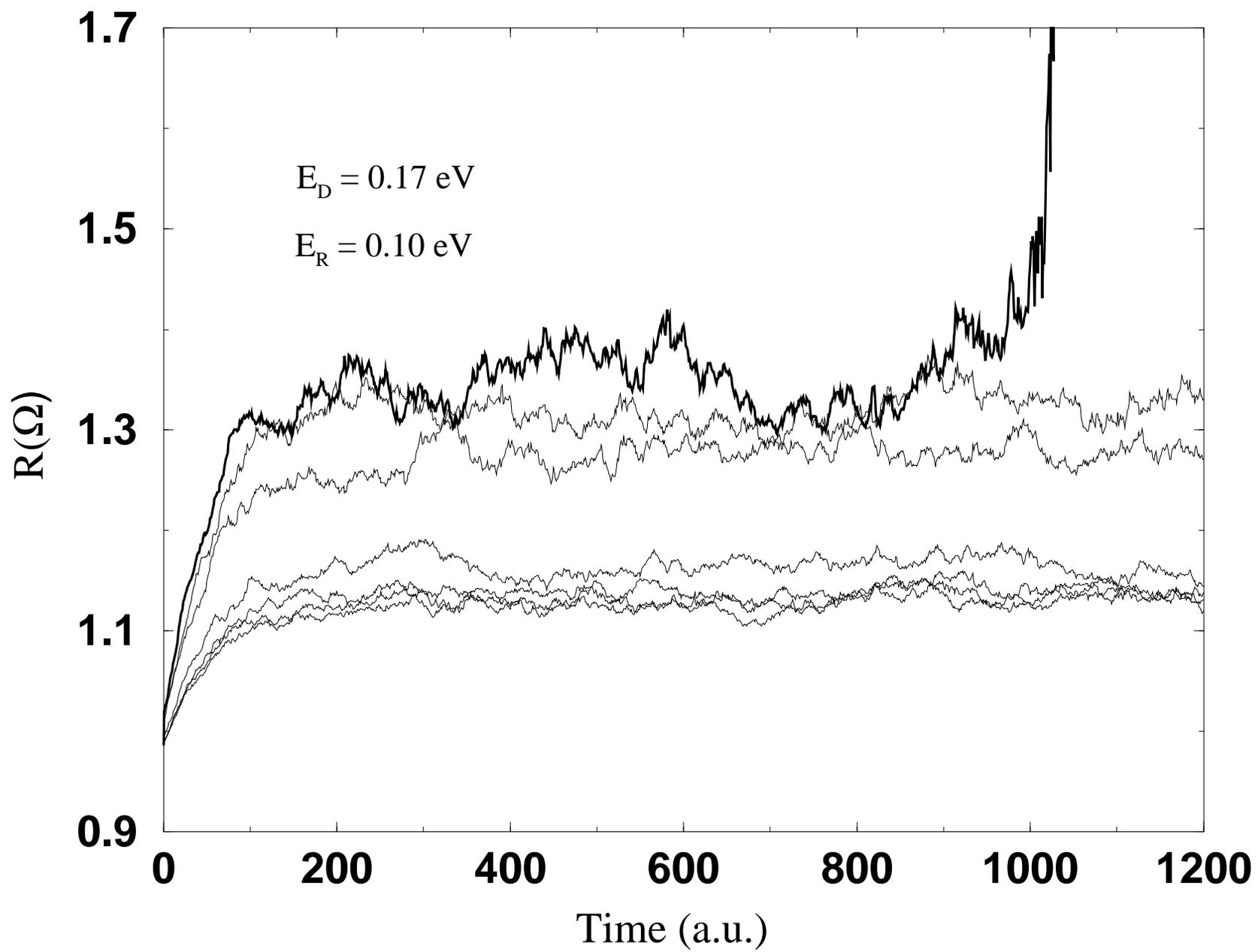
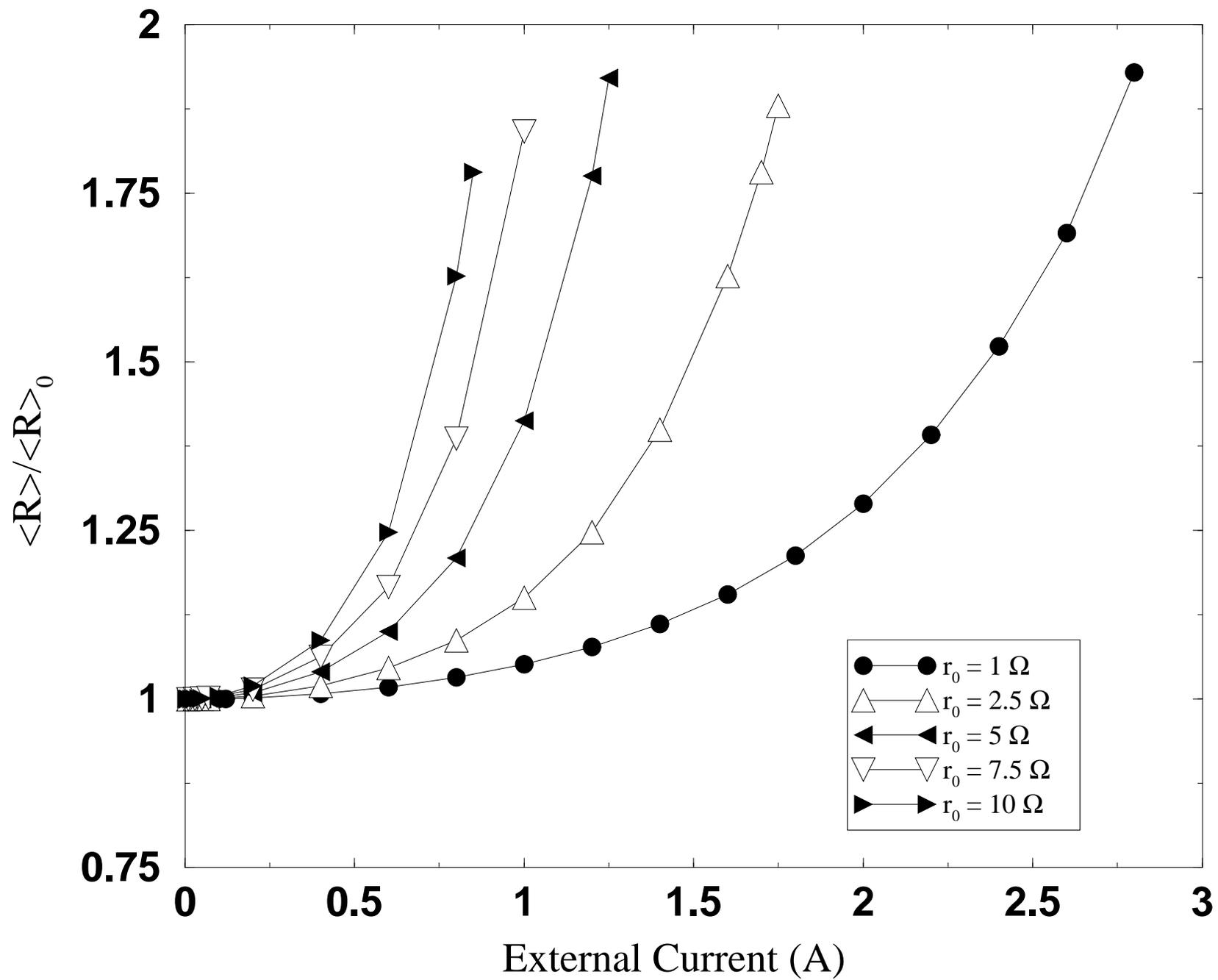
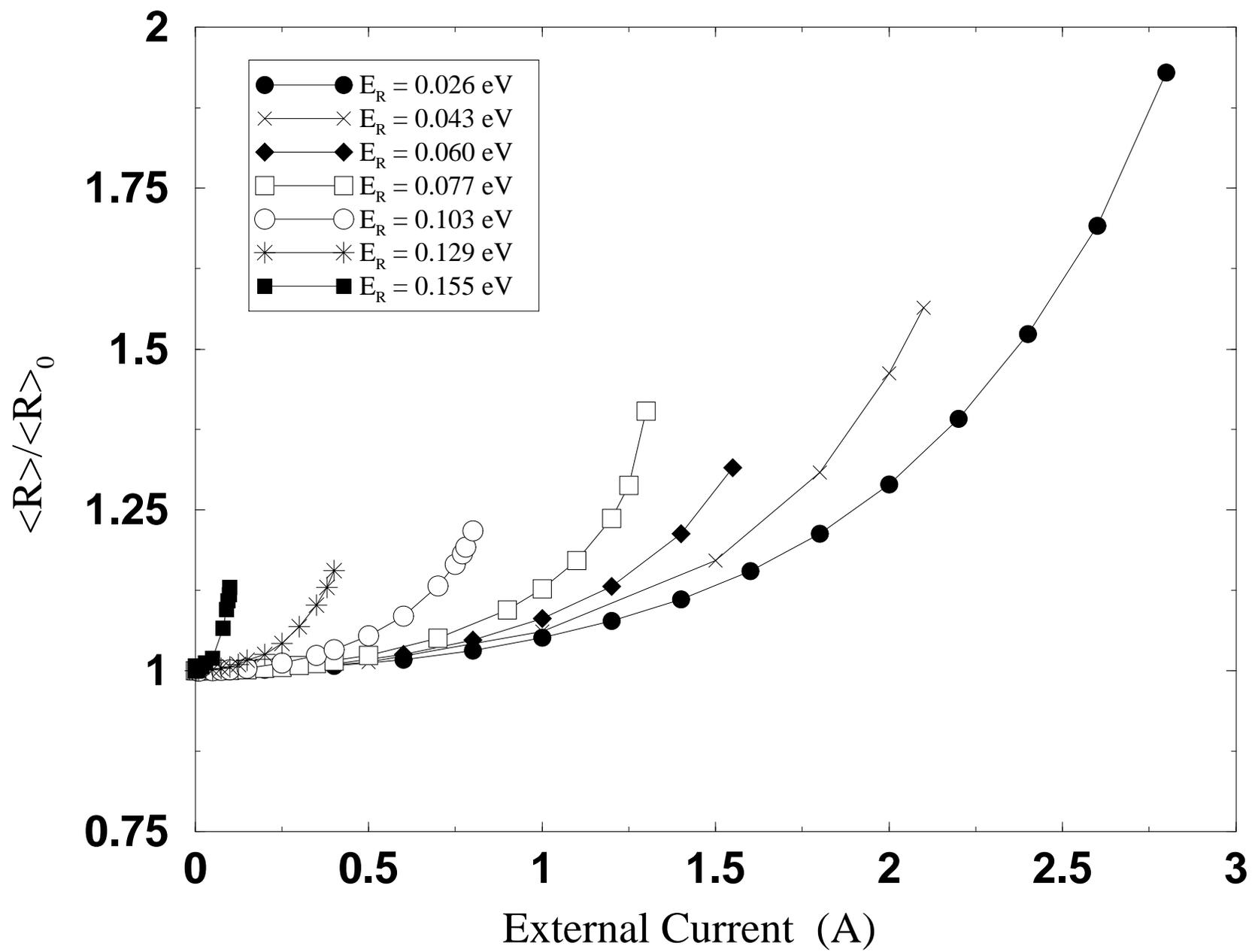
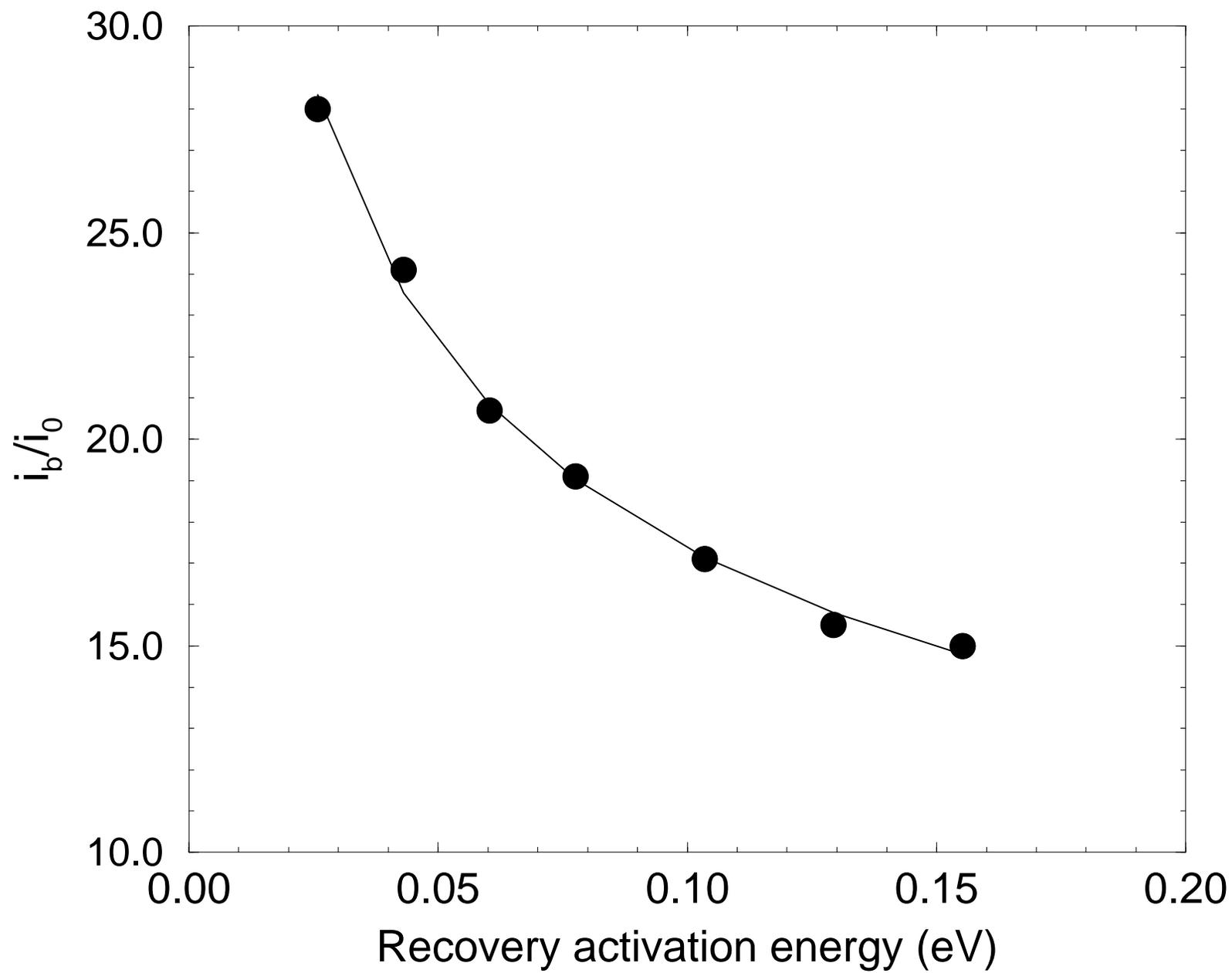
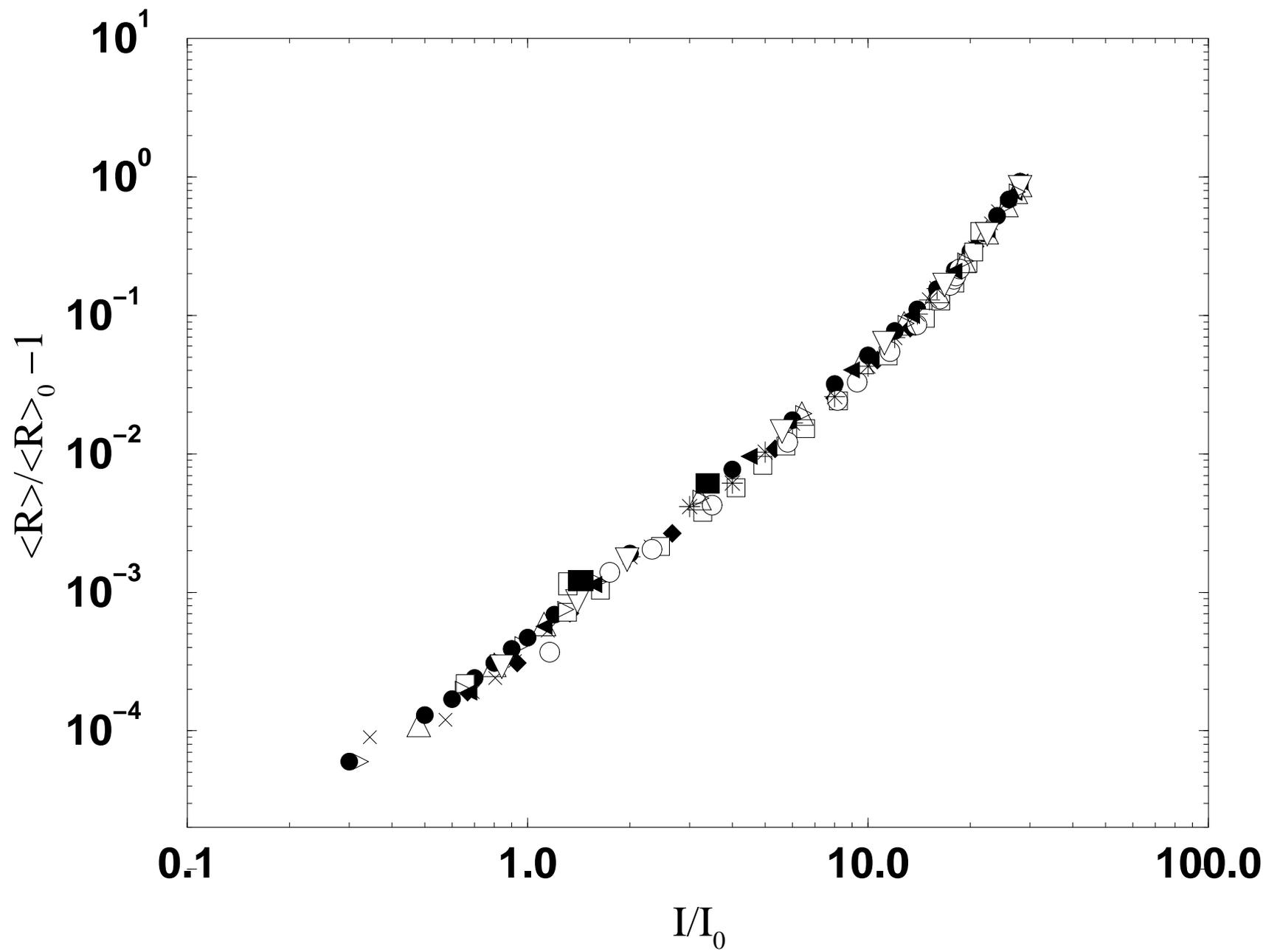
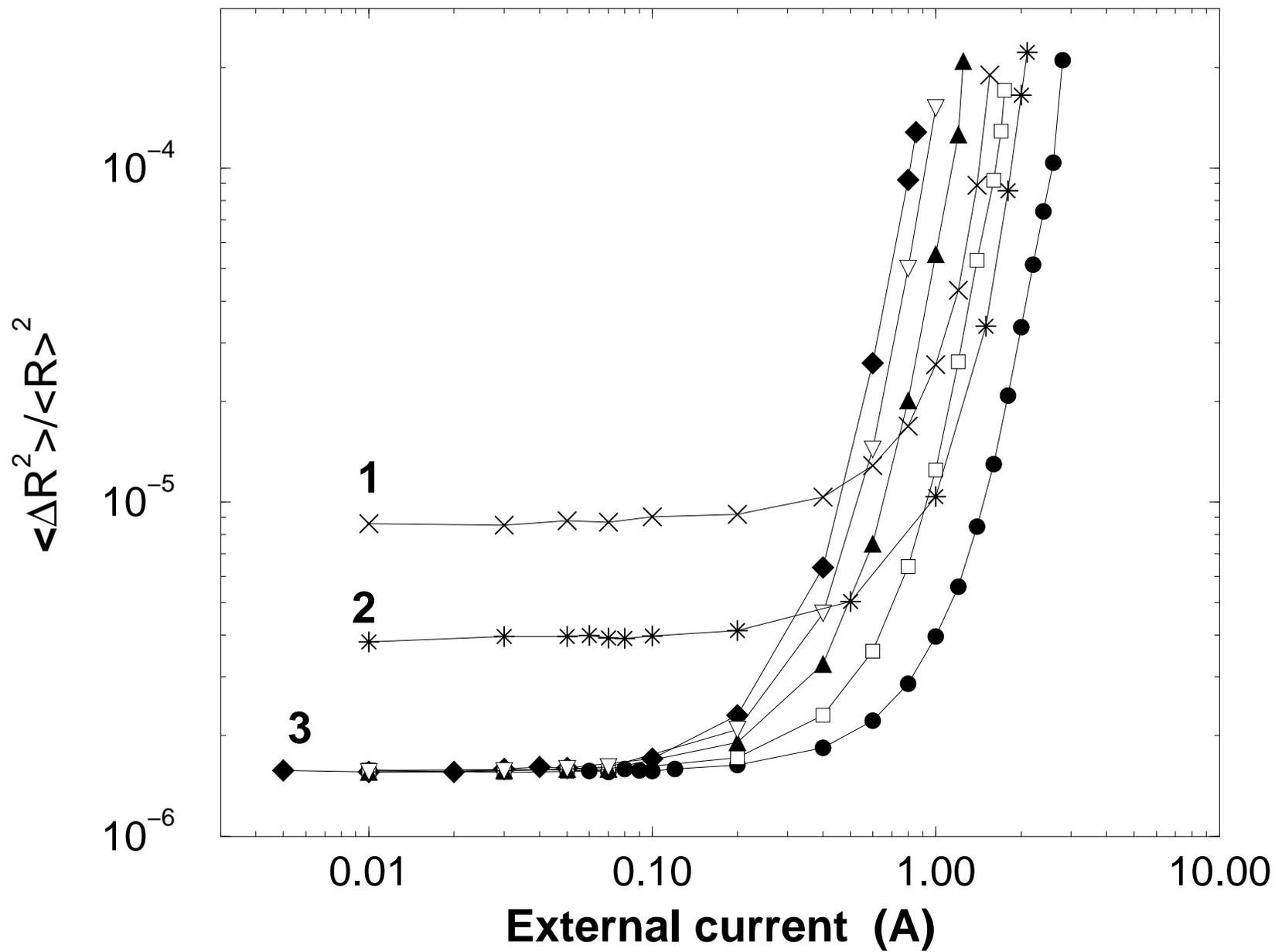
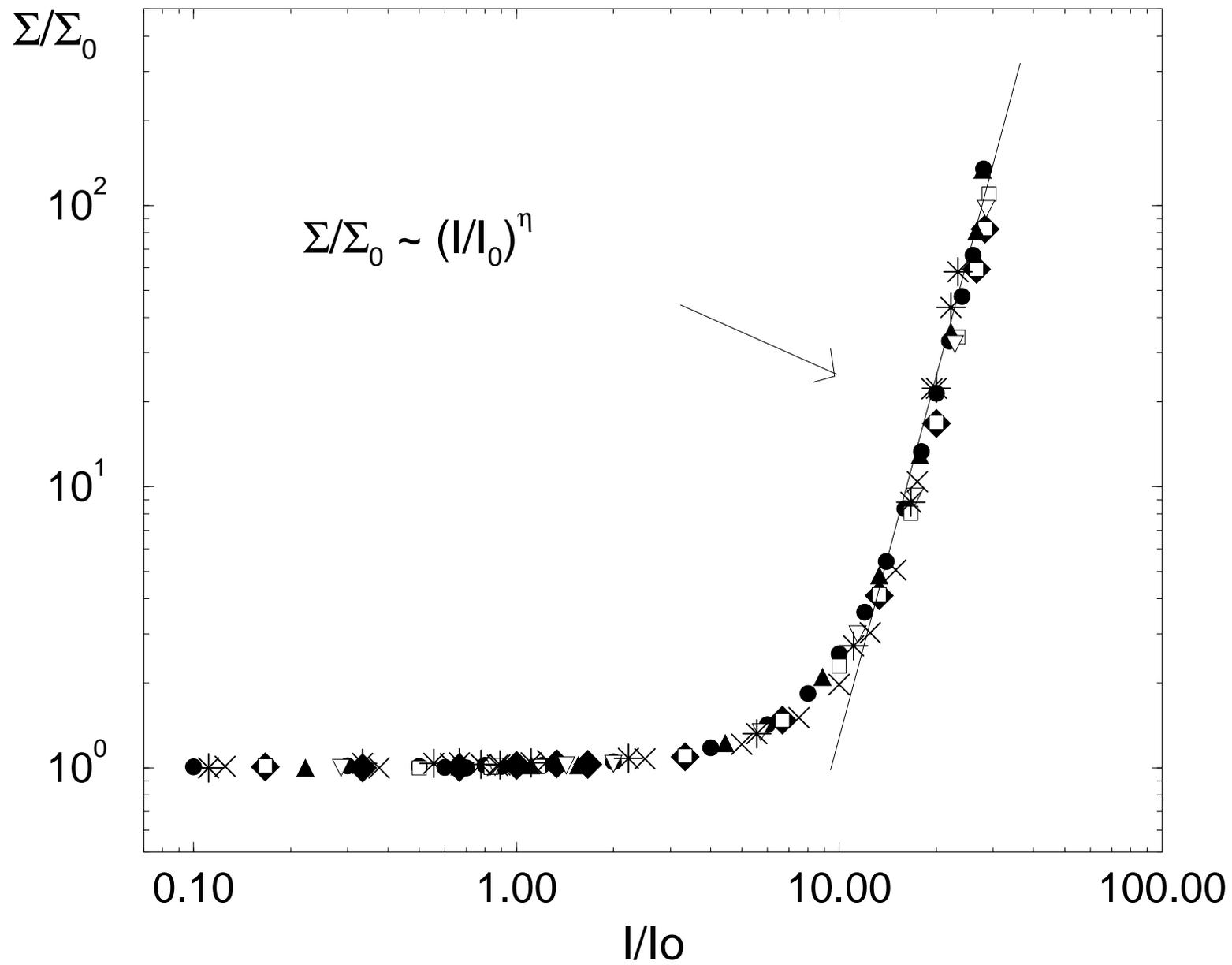
\end{document}